\documentclass[aps,prb,twocolumn,superscriptaddress,showpacs]{revtex4-1}
\usepackage{graphicx}
\usepackage{amssymb}
\usepackage{amsmath}
\usepackage{appendix}
\usepackage{comment}

\begin{document}

\title{Dynamic phase diagram of dc-pumped magnon condensates}

\author{Scott A. Bender}
\affiliation{Department of Physics and Astronomy, University of California, Los Angeles, California 90095, USA}

\author{Rembert A. Duine}
\affiliation{Institute for Theoretical Physics and Center for Extreme Matter and Emergent Phenomena, Utrecht University, Leuvenlaan 4, 3584 CE Utrecht, The Netherlands}

\author{Arne Brataas}
\affiliation{Department of Physics, Norwegian University of Science and Technology, NO-7491 Trondheim, Norway}

\author{Yaroslav Tserkovnyak}
\affiliation{Department of Physics and Astronomy, University of California, Los Angeles, California 90095, USA}

\begin{abstract}
We study the effects of nonlinear dynamics and damping by phonons on a system of interacting electronically pumped magnons in a ferromagnet. The nonlinear effects are crucial for constructing the dynamic phase diagram, which describes how ``swasing" and Bose-Einstein condensation emerge out of the quasiequilibrated thermal cloud of magnons. We analyze the system in the presence of magnon damping and interactions, demonstrating the continuous onset of stable condensates as well as hysteretic transitions.
\end{abstract}

\maketitle

\section{Introduction}

Elementary excitations of uniform ferromagnets (magnons) are bosonic in nature, thus exhibiting properties similar in character to those of cold atoms, photons, and excitons.  Each of these systems can undergo a bosonic condensation wherein the lowest-energy mode displays a macroscopic occupation. The condensate, thereafter, manifests a macroscopic phase, spontaneously breaking U(1) gauge symmetry. Magnons have been expected \cite{kalafatiJETPL89,*kalafatiJETP91} and observed \cite{demokritovNAT06} to undergo condensation under microwave pumping. Their condensate phase has a transparent physical interpretation as the precessional angle of collective magnetic dynamics.

In Ref.~\onlinecite{benderPRL12}, magnon condensates are proposed to be realized through dc electronic pumping. To this end, a ferromagnetic insulator, e.g., yttrium iron garnet (YIG), is directly attached to a conducting normal metal. Spin-pumping by the precessing magnet (or spin waves), governed by a sizable spin-mixing conductance across the interface, results in a loss of magnons and the corresponding creation of particle-hole excitations in the normal metal.  This magnetic bleeding may be overcome either by increasing the current in the normal metal, which transports angular momentum into the ferromagnet by the spin Hall effect,\cite{andoPRL08,*liuPRL11sh} or by utilizing a temperature gradient across the interface, thus actuating the spin Seebeck effect.\cite{uchidaNAT08,*uchidaNATM10} Under a critical spin Hall and/or Seebeck biases, an excess of incoherently pumped magnons can precipitate a spontaneous condensation.

\begin{figure}[pth]
\includegraphics[width=\linewidth,clip=]{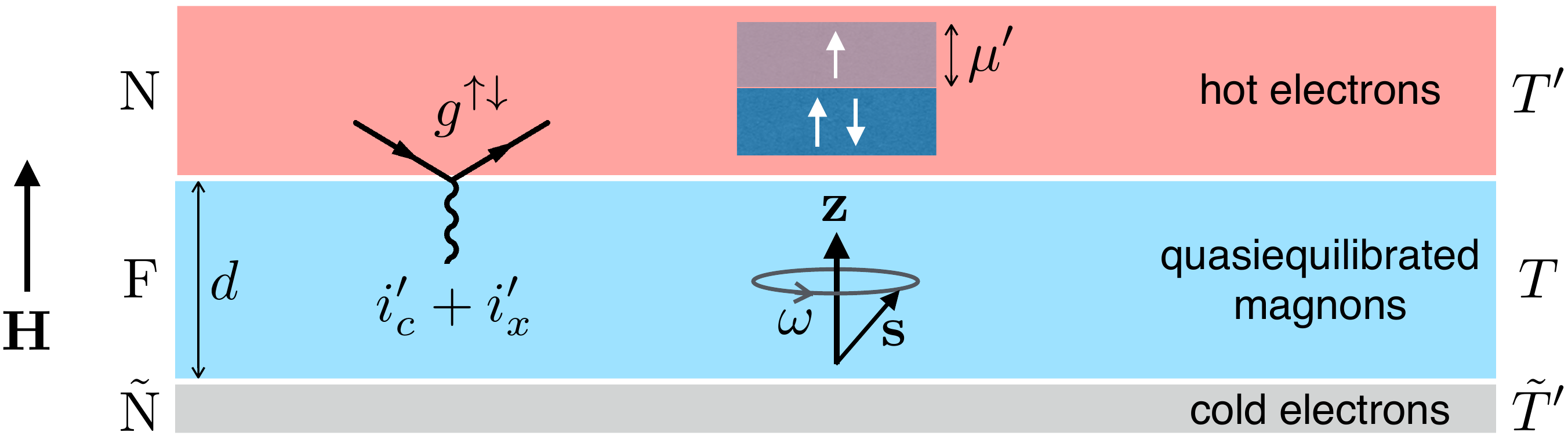}
\caption{Schematic of the proposed heterostructure. On the top, the normal metal N, with electron temperature $T'$, provides spin torque through spin accumulation $\mu'$ in the $z$ direction at the interface with the ferromagnet (F). The F is assumed to be sufficiently thin such that its magnon temperature $T$ is uniform throughout. When the equilibrium spin density in the F points in the $-z$ direction, spin accumulation $\mu'>0$ (along $z$) can supply magnons. The normal metal $\tilde{\rm N}$ is a poor spin sink which can, nevertheless, exchange energy with magnons and phonons in the ferromagnet. The phonon temperature in the F is thus in general determined by the normal-metal electron temperatures $T'$ and $\tilde{T}'$, given by their average $(T'+\tilde{T}')/2$ in the symmetric case.}
\label{schematic}
\end{figure}

In this paper, we build upon the proposal in Ref.~\onlinecite{benderPRL12} by making two important advancements. First, we include nonlinear effects associated with magnon-magnon interactions in the presence of finite-angle collective precession. Nonlinear effects can play an important role in stabilizing coherent dynamics under large spin Hall/Seebeck biases, as well as accounting for the interaction of the condensate with the thermal magnon cloud. Second, we include Gilbert damping due to magnon-lattice coupling and allow for an additional energy-sink channel by attaching a poor spin-sink normal metal on the other side of the ferromagnet. See Fig.~\ref{schematic} for a schematic of our setup. The role of this second normal metal in our model is to (i) anchor the adjacent lattice temperature and (ii) provide a reservoir that dissipates excess energy injected along with magnons from the first normal metal, which helps in fomenting condensation.

The paper is organized as follows. In Sec.~\ref{md}, we start by constructing the nonlinear dynamics of the condensate (A) and incoherent dynamics of the cloud (B), including its interaction with the condensate. In Sec.~\ref{tre}, we derive rate equations for spin and energy transfer into the normal metal N and the phonon bath. In Sec.~\ref{dpd}, the dynamic phase diagram of the pumped magnetic system is constructed, focusing on two special limits: (A) the fixed magnon temperature regime, which is controlled by spin flows between different subsystems (of magnons, electrons, and phonons), (B) the floating magnon temperature regime, in which the steady state is determined by self-consistent flows of both spin and energy. In both cases, we find regions of stable condensate with second-order as well as first-order hysteretic transitions out of the normal phase. Finally, Sec.~\ref{so} summarizes our findings and offers an outlook.

\section{Magnetic dynamics}
\label{md}

\subsection{Condensate dynamics}
\label{cd}

We start by considering dynamics at absolute zero temperature, assuming only the lowest mode is excited. For simplicity, we neglect magnetostatic effects, such that the lowest-frequency excitation is given by homogeneous (monodomain) magnetic precession. Supposing, furthermore, cylindrical symmetry about the $z$ axis, the effective monodomain Hamiltonian can be phenomenologically expanded as
\begin{equation}
\mathcal{H}=HS_z+\frac{KS_z^2}{2S}\,,
\label{H0}
\end{equation}
where $S$ is the total (macro)spin of the ferromagnet (in units of $\hbar$), $S_z$ is its $z$-axis projection, $H$ is the applied field in the $z$ direction (upon absorbing the gyromagnetic ratio), and $K$ is the axial anisotropy (with $K>0$ corresponding to an easy $xy$ plane). We suppose that $H>K$, such that spin is oriented in the $-z$ direction in the ground state.

The spin algebra, $[S_i,S_j]=i\epsilon_{ijk}S_k$, can be conveniently recast in terms of the Holstein-Primakoff bosons:\cite{holsteinPR40}
\begin{equation}
S_+=a^\dagger\sqrt{2S-a^\dagger a}\,,\,\,\,S_z=a^\dagger a-S\,,
\end{equation}
where $S_+\equiv S_x+iS_y$, $S$ is the total spin, and $a$ is a time-dependent ground-state magnon operator satisfying commutation relation $[a,a^\dagger]=1$.  The Hamiltonian in Eq.~(\ref{H0}) is thus rewritten in terms of free-boson and interacting contributions (dropping a constant offset):
\begin{equation}
\mathcal{H}=(H-K)a^\dagger a+\frac{K(a^\dagger a)^2}{2S}\,.
\end{equation}
A classical precession for large spin $S$ corresponds to a coherent state for boson $a$: $|\alpha\rangle=e^{\alpha a^\dagger-\alpha^*a}|0\rangle$, such that $a|\alpha\rangle=\alpha|\alpha\rangle$, where $|0\rangle$ is the ground state with $S_z=-S$. The quantum-to-classical correspondence is provided by $\alpha^*\sqrt{2S-|\alpha|^2}\leftrightarrow S_+$, where the phase of $\alpha=|\alpha|e^{-i\phi}$ corresponds to the azimuthal angle of spin $\mathbf{S}$ in the $xy$ plane: $\phi=\tan^{-1}(S_x/S_y)$. For small-angle precession, $|\alpha|^2\ll S$:
\begin{equation}
S_+\approx\alpha^\ast\sqrt{2S}=\sqrt{2SN}e^{i\phi}\,,
\label{S}
\end{equation}
where $N=|\alpha|^2=S_z+S$.

In the Heisenberg picture,
\begin{align}
i\hbar\partial_ta&=-[\mathcal{H},a]=\partial_{a^\dagger}\mathcal{H}\nonumber\\
&=(H-K)a+\frac{K\{a^\dagger a,a\}}{2S}\to\hbar\omega a\,,
\end{align}
where $\hbar\omega=H-K+K(N-1/2)/S$, when acting on the magnon-number, $N\equiv a^\dagger a$, eigenstate $|N\rangle$, and $\{,\}$ stands for the anticommutator. (This $\hbar\omega$ corresponds to the energy for adding a magnon to the state $|N-1\rangle$.) In the classical limit, $S\gg1$, this gives the familiar Larmor precession frequency:
\begin{equation}
\hbar\omega\equiv\hbar\dot{\phi}=\partial_N\mathcal{H}=H+\frac{KS_z}{S}\,.
\label{w}
\end{equation}
Indeed, the variables $\hbar S_z$ and $\phi$ are canonically conjugate: $\hbar\dot{\phi}=\partial_{S_z}\mathcal{H}$, $\hbar\dot{S_z}=-\partial_{\phi}\mathcal{H}$. Viewing this as a special (cylindrically-symmetric) instance of the Landau-Lifshitz equation,\cite{landauBOOKv9} we can easily extend the Hamiltonian \eqref{H0} to include more general magnetic interactions. A common phenomenology for dissipation, furthermore, is provided by the Gilbert damping,\cite{gilbertIEEEM04} which endows frequency \eqref{w} with an imaginary component, $\omega\to\omega(1-i\alpha)$, where $\alpha$ is a material-dependent constant. The corresponding magnon-number relaxation rate, $\tau^{-1}=2\alpha\omega$, is proportional to the precession frequency.

\subsection{Thermal cloud}
\label{tc}

At a finite temperature $T$, the thermally-excited magnons also contribute to the total spin angular momentum. For large bulk samples of volume $V$, it is now natural to switch from the total spin $\mathbf{S}$ to the spin density $\mathbf{s}=\delta\mathbf{S}/\delta V$. Extending Eq.~\eqref{S} to this case, while assuming that $T\ll T_c$, the Curie temperature (such that we limit our attention to small-angle magnetic dynamics), this spin density can be written [relative to the saturated value $-s\mathbf{z}$ at $T=0$, where $s=S/V$]:
\begin{equation}
\mathbf{s}\approx\left(\sqrt{2s}\Re\psi,\sqrt{2s}\Im\psi, n \right)\,.
\label{ss}
\end{equation}
Here, $n=n_c+n_x$, in terms of the condensate magnon density $n_c$ (i.e., density of magnons occupying the lowest mode) and the thermal cloud density $n_x$ (which is composed of the magnon states excited above the lowest-energy mode); $\psi\equiv\sqrt{n_c} e^{i\phi}$ plays the role of the condensate order parameter, with $\phi$ being the $xy$-plane azimuthal angle of the coherent spin precession. Note that only the magnon condensate component contributes to the $xy$ spin-density projections.

The intrinsic dynamics of magnons with wavenumber $q$, by extension of Eq.~\eqref{w}, is given by
\begin{equation}
\hbar\omega_q=H-K\left(1-\frac{n_c}{s}\right)+Aq^2=\hbar\Omega+K\frac{n_c}{s}+Aq^2\,,
\label{wq}
\end{equation}
where $A$ is the ferromagnetic exchange stiffness (in appropriate units) and $\Omega\equiv(H-K)/\hbar>0$ is the (monodomain) ferromagnetic-resonance frequency. Here, for simplicity, we are retaining only the nonlinear term stemming from the anisotropy term $KS_z^2/2S$ in the Hamiltonian,\cite{Note1} which would arise from, e.g., the global shape anisotropy.\cite{suhlJPCS57} This is justified so long as the key nonlinearity stems from the feedback of the condensate $n_c$ on the frequency of the magnon modes.\cite{Note2} Gilbert damping still gives $\tau_q^{-1}=2\alpha\omega_q$ for the $q$-dependent relaxation rate.

\section{Transport Rate Equations}
\label{tre}

The rate equation for the magnon-number density, $\dot{n}=\dot{n}_c+\dot{n}_x$, is governed by the Landau-Lifshitz-Gilbert (LLG) dynamics of the condensed and thermal magnons, including their interactions, damping of spin and energy to the lattice, and spin and energy transport between the ferromagnet and the normal-metal reservoirs that are governed by the electron-magnon scattering. The zero-temperature condensate dynamics are described by the classical LLG equation of motion (extended to include spin-transfer torques) for the unit-vector collective spin direction $\mathbf{n}$:
\begin{equation}
(1+\alpha\mathbf{n}\times)\hbar\dot{\mathbf{n}}+\mathbf{n}\times\mathbf{H}_{\rm eff}=\left(\Im\alpha'+\Re\alpha'\mathbf{n}\times\right)\left(\boldsymbol{\mu}'\times\mathbf{n}-\hbar\dot{\mathbf{n}}\right)\,,
\label{llgs}
\end{equation}
where $\mathbf{H}_{\rm eff}\equiv\partial_\mathbf{S}H=(H+K\mathbf{n}\cdot\mathbf{z})\mathbf{z}$ is the effective field, $\boldsymbol{\mu}'=\mu'\mathbf{z}$ is the vectorial spin accumulation in N,
\begin{equation}
\alpha'=\Re\alpha'+i\Im\alpha'\equiv\frac{g^{\uparrow \downarrow}}{4\pi s d}\,,
\end{equation}
in terms of the complex-valued spin-mixing conductance $g^{\uparrow \downarrow}$ (in units of $e^2/h$, and per unit area) of the F$\mid$N interface and the F layer thickness $d$. The left-hand side of Eq.~\eqref{llgs} is the standard LLG equation,\cite{landauBOOKv9,*gilbertIEEEM04} while the right-hand side consists of the static spin-transfer torques\cite{slonczewskiJMMM96,bergerPRB96} $\propto\boldsymbol{\mu'}$ and spin-pumping torques\cite{tserkovPRL02sp,*tserkovRMP05} $\propto\dot{\mathbf{n}}$ (which are Onsager reciprocal\cite{tserkovPRB08mt}).

Rewriting Eq.~\eqref{llgs} in spherical coordinates, in terms of the condensate density $n_c$,
\begin{equation}
\mathbf{n}=(n_\perp\cos\phi,n_\perp\sin\phi,n_c/s-1)\,,
\end{equation}
where $n_\perp=\sqrt{2n_c/s-(n_c/s)^2}$, we have
\begin{equation}\begin{aligned}
\left(1+\Im\alpha'\right)\hbar\dot{n}_c&=-\left[\left(\alpha+\Re\alpha'\right)\hbar\dot{\phi}-\Re\alpha'\mu'\right](2n_c-n_c^2/s)\,,\\
\left(1+\Im\alpha'\right)\hbar\dot{\phi}&=\hbar\omega+\Im\alpha'\mu'+\left(\alpha+\Re\alpha'\right)\frac{\hbar\dot{n}_c}{2n_c-n_c^2/s}\,.
\label{nandphi}
\end{aligned}\end{equation}
Here, $\omega\equiv\omega_0$ is given by Eq.~\eqref{wq}, with $q=0$. These equations generalize the Hamilton's equations of motion for the canonically-conjugate pair of variables $(n_c,\phi)$ to include dissipation (magnon-lattice coupling) and spin-transfer torques/spin-pumping (magnon-electron coupling). Assuming that $\alpha,|\alpha'|\ll1$, which is nearly always the case in practice, Eqs.~\eqref{nandphi} give for the condensate rate equation 
\begin{equation}
\hbar\dot{n}_c=i_c+i'_c\,,
\label{ncc}
\end{equation}
where
\begin{equation}
i_c+i'_c=-2\left[(\alpha+\Re\alpha')\hbar\omega-\Re\alpha'\mu'\right]n_c(1-n_c/2s)
\label{icc}
\end{equation}
captures the effects of Gilbert damping and spin-transfer torque. [Here, we combined the expressions for $\dot{n}_c$ and $\dot{\phi}$ in Eq.~(\ref{nandphi}) and dropped the terms that are quadratic in $\alpha$ and $\alpha'$.] Since $\Im\alpha'$ is eliminated by this substitution, hereafter $\alpha'$ stands for $\Re\alpha'$ only.

According to Eq.~\eqref{icc}, the condensate rate of change \eqref{ncc} is scaled by the geometrical factor $1-n_c/2s$, which can be divided out and, if $n_c\ll s$, disregarded. When in the following we complement the magnon rate equation with thermal contributions, this factor could be absorbed by an appropriate rescaling of the thermal terms, which would lead to small cross terms between the quantities associated with the condensate and the thermal cloud. The Gilbert-damping and spin-transfer contributions to the zero-temperature condensate rate equation are then respectively given by
\begin{align}
\label{ic}
i_c&=-2\alpha\hbar\omega n_c\,,\\
i'_c&=-2\Re\alpha'(\hbar\omega-\mu')n_c\,.
\label{icp}
\end{align}
Equation \eqref{icp} was derived in Ref.~\onlinecite{benderPRL12} in a perturbative treatment of the electron-magnon scattering, which is consistent with neglecting terms that are quadratic in $\alpha$'s.

At finite temperatures,
\begin{equation}
\hbar\dot{n}_c=(i_c+i'_c)+i_{xc}\,,
\label{ncdot}
\end{equation} where $i_{xc}$ is the rate of spin transfer from the thermal cloud to condensate. The thermally-excited magnons also obey generalized LLG/spin-torque relations, which we derive below. In order to simplify the following discussion, we will limit our attention to the situations when spin-preserving magnon-magnon interactions are fast enough that magnons form a Bose-Einstein distribution with an effective temperature $T=(k_B\beta)^{-1}$ and chemical potential $\mu$. The total thermal-cloud density is then given by
\begin{equation}
n_x=\int_0^\infty d\epsilon D(\epsilon)n_{\rm BE}\left[\beta\left(\epsilon-\mu^\ast\right)\right]\,,
\end{equation}
where $\mu^\ast\equiv\mu-\hbar\omega\leq0$ is the magnon chemical potential relative to the band edge (set at $\epsilon=0$), which, on the absolute scale, is shifted by the condensate frequency $\hbar\omega$; $D(\epsilon)=\sqrt{\epsilon}/4\pi^2A^{3/2}$ is the magnon density of states; and $n_{\rm BE}(x)\equiv(e^x-1)^{-1}$. Writing the thermal-cloud rate equation, $\hbar\dot{n}_x=i_x+i'_x$, in terms of the Gilbert-damping, $i_x$, and spin-torque, $i_x'$, contributions, we assert for the former:
\begin{equation}
i_x=\hbar\int_0^\infty d\epsilon D(\epsilon)\frac{n_{\rm BE} \left[\beta'' \left(\epsilon+\hbar \omega \right) \right]-n_{\rm BE} \left[\beta \left(\epsilon-\mu^\ast\right)\right]}{\tau(\epsilon)}\,,
\label{in}
\end{equation}
where $\beta''\equiv(k_BT'')^{-1}$ is the inverse (effective) temperature of phonons (which are assumed to be responsible for the Gilbert damping) and $\hbar/\tau(\epsilon)=2\alpha(\hbar\omega+\epsilon)$. The spin-torque rate is given by\cite{benderPRL12}
\begin{align}
\label{inp}
i_x'=&4 \alpha' \int_0^\infty d\epsilon D(\epsilon)(\epsilon+\hbar\omega-\mu')\\ 
&\times \left\{n_{\rm BE} \left[\beta' \left(\epsilon+\hbar\omega-\mu'\right) \right] -n_{\rm BE} \left[\beta \left(\epsilon-\mu^* \right)\right] \right\}\,,\nonumber
\end{align}
where $\beta'$ is the inverse normal-metal N electron temperature.\cite{Note} The rate equation for the thermal-cloud is given by
\begin{equation}
\hbar \dot{n}_x=(i_x+i_x')+i_{cx}\,,
\label{nxdot}
\end{equation}
where $i_{cx}$ is the rate of spin transfer from the condensate to cloud. 

The total spin current $i$ passing through the normal-metal interface is found by adding Eqs.~(\ref{ncdot}) and~(\ref{nxdot}):
\begin{equation}
i=\hbar \dot{n}_c+\hbar \dot{n}_x=(i_c+i'_c)+(i_x+i'_x)\,,
\label{totsc}
\end{equation}
where we set $i_{xc}+i_{cx}=0$, assuming magnon-number preserving magnon-magnon interactions (which is rooted in spin conservation for a cylindrically-symmetric magnetic system). The expression for the net spin current $i$, using rate equations for the condensate, Eqs.~\eqref{ic} and \eqref{icp}, and thermal cloud, Eqs.~\eqref{in} and \eqref{inp}, forms one of our key results. In order to find steady states, we will have to solve for $i=0$. Subject to external conditions of pumping, two unknowns thus need to be established: the effective temperature, $T$, and chemical potential, $\mu$, of magnons.

In order to evaluate a common temperature and chemical potential for the magnons, we also need to consider the energy flow into the system. The total magnon energy density in our model is given by $e=e_c+e_x$, where
\begin{equation}
e_c=\hbar\Omega n_c+\frac{Kn_c^2}{2s}
\end{equation}
is the condensate energy and
\begin{equation}
e_x=\int_0^\infty d\epsilon(\epsilon+\hbar\omega)D(\epsilon) n_{\mathrm{BE}}\left[ \beta (\epsilon-\mu^*) \right]
\end{equation}
is the thermal-cloud energy. We recall, in particular, that $\omega\equiv\omega_0=\Omega+Kn_c/\hbar s$ in the above equations is affected by the presence of the condensate $n_c$. The total energy-transfer rate from N and the lattice into magnons is thus given by:
\begin{equation}
j=\dot{e}_c+\dot{e}_x=\left(\omega+\frac{K}{\hbar s}n_x\right)\left(i_c+i'_c\right)+(j_x+j'_x)\,,
\label{ecrate}
\end{equation}
where $j_x$ and $j_x'$ are given by the expressions similar to Eqs.~\eqref{in} and \eqref{inp} but with an additional factor of $(\omega+\epsilon/\hbar)$ in the integrands.\cite{Note3}

\section{Dynamic phase diagrams}
\label{dpd}

\subsection{Fixed magnon temperature}
\label{fmt}

If we suppose that the magnetic coupling to the phonons is sufficiently strong that the magnon temperature $T$ is fixed by the phonon reservoir, $T\to T''$, we may disregard the energy current $j$. In this limit, the spin current $i$ fully determines the state of the system. The magnon temperature in a magnetic film sandwiched by two metals, as sketched in Fig.~\ref{schematic}, can similarly be fixed by the electron temperatures $T'$ and $\tilde{T}'$ [for example, $T\to(T'+\tilde{T}')/2$ in a mirror-symmetric structure], either through direct magnon-electron scattering at the interfaces or via magnon-phonon interaction. Under the reigning assumption that the magnonic cloud and condensate maintain internal thermodynamic equilibrium at all times, the magnet is always either in normal phase (NP) or condensate phase (CP). Then, only one variable is left free to vary: $\mu^*$ in NP or $n_c$ in CP, which is controlled by the spin current $i$ flowing into the magnetic subsystem.

In a normal phase, the condensate is absent ($n_c=0$), and the magnon current goes entirely into the thermal cloud:
\begin{equation}
\hbar \dot{n}_x=i\,,
\end{equation}
where $i=i_x+i_x'$ consists only of the normal component, Eqs.~\eqref{in}, \eqref{inp}, which depend on $\mu^*$. We will be treating the dependence $i(\mu^*)$ inside NP numerically.

If the magnons are condensed (i.e., $\mu^*=0$) while their temperature $T$ is fixed, the spin current, Eq.~\eqref{totsc}, must, via magnon-magnon interactions, be entirely transformed into the condensate density:
\begin{equation}
\hbar \dot{n}_c=i\,.
\label{dnc}
\end{equation}
Even in this simple limit, however, we cannot obtain an exact analytic solution for $n_c(t)$, since the flux $i$ has an implicit nonlinear dependence on $n_c$ [through the dependence of $(i_x+i'_x)$ on $\omega(n_c)$]. When $n_c/s\ll1$, which is the limit we are focusing on throughout, we can expand $i$ in its powers:
\begin{equation}
i=\imath_x-\sigma\frac{n_c}{s}-\zeta\left(\frac{n_c}{s}\right)^2+\mathcal{O}\left(\frac{n_c}{s}\right)^3\,.
\label{ia}
\end{equation}
Here, $\imath_x\equiv i_x+i_x'$, after setting $\mu^*=0$ and $n_c=0$ in Eqs.~\eqref{in} and \eqref{inp}. According to Eqs.~\eqref{ic} and \eqref{icp},
\begin{equation}
\sigma=2s(\alpha+\alpha')\hbar\Omega-2s\alpha'\mu'+\delta\sigma
\label{s}
\end{equation}
and
\begin{equation}
\zeta=2s(\alpha+\alpha')K+\delta\zeta\,,
\label{z}
\end{equation}
where $\delta\sigma$ and $\delta\zeta$ are thermal-magnon corrections. Using Eqs.~\eqref{in} and \eqref{inp}, the latter are evaluated at $k_BT\gg\hbar\Omega$ to be:
\begin{equation}\frac{
\delta\sigma}{s K}\sim(\alpha+2\alpha')\left(\frac{T}{T_c}\right)^{3/2}
\end{equation}
and
\begin{equation}
\frac{\delta\zeta}{sK}\sim-(\alpha+2\alpha')\sqrt{\frac{T}{T_c}}\frac{K}{k_BT_c}\,,
\end{equation}
up to numerical factors of order unity. Here, $k_B T_c\sim s^{2/3} A$ is the Curie temperature. These corrections are clearly unimportant, so long as $K\ll k_BT_c$ (recalling that $T\ll T_c$ throughout), and will be omitted in the following. We thus conclude, in particular, that $\zeta>0$.

\subsubsection{Swasing}

We start by considering the low-temperature limit of a stiff ferromagnet, where the thermal-current contribution $\imath_x$ in Eq.~\eqref{ia} can be disregarded. The condensate dynamics, $\hbar\dot{n}_c=i$, is then governed by two transport coefficients: $\sigma$ and $\zeta$.

The coefficient $\sigma$ in Eq.~(\ref{ia}) represents an effective damping of the condensate and describes a competition between, on the one hand, damping by phonons and electrons (captured by the first term in $\sigma$, proportional to $\Omega>0$, where $\alpha$ parametrizes Gilbert damping and $\alpha'$ spin pumping\cite{tserkovPRL02sp}) and, on the other, spin-transfer torque from the normal-metal N (captured by second term in $\sigma$, proportional to spin accumulation $\mu'$).  When the former contribution is larger, $\sigma$ is positive, and the torque provided by the second term Eq.~(\ref{ia}) relaxes the condensate spin density (with the total spin decaying towards the $-z$ axis). Conversely, upon the application of a sufficiently large and positive spin accumulation $\mu'$, $\sigma$ is negative, and the net torque from the linear in $n_c$ term in Eq.~(\ref{ia}) drives the condensate spin towards the $+z$ axis. The quadratic term proportional to $\zeta$ in Eq.~\eqref{ia} describes a nonlinear enhancement of damping, which ultimately curbs the exponential growth of the condensate when $\sigma<0$, leading to the fixed point
\begin{equation}
\frac{n_c}{s}\to\frac{|\sigma|}{\zeta}=\frac{\left| (1+\alpha/\alpha')\hbar \Omega -\mu' \right|}{(1+\alpha/\alpha') K}\,.
\end{equation}

In the absence of intrinsic Gilbert damping, i.e., $\alpha=0$, the effective damping $\sigma$ is proportional to $\hbar\Omega-\mu'$. This was first pointed out by Berger,\cite{bergerPRB96} who coined the term \textit{swaser} (spin-wave amplification by stimulated emission of radiation) to describe the coherent emission of spin waves, signified by negative damping, when the pumping $\mu'$ overcomes the intrinsic threshold associated with the gap $\hbar\Omega$. This \textit{swasing} instability may be understood thermodynamically: Because the condensate carries no entropy, the free-energy change due the creation of $\delta N$ magnons and the corresponding annihilation of the up-electron/down-hole pairs is
\begin{equation}
\delta F=(\hbar\Omega-\mu')\delta N\,.
\end{equation}
When $\mu'<\hbar \Omega$, the condensate is damped by the transfer of angular momentum and energy out of the magnet into N; when $\mu'>\hbar \Omega$, however, the absorption of energy by the condensate becomes entropically beneficial, signaling an instability.

A finite $\alpha$ in Eq.~\eqref{s} raises the swasing instability threshold to
\begin{equation}
\mu'=\left(1+\frac{\alpha}{\alpha'}\right)\hbar\Omega\,,
\label{sw}
\end{equation}
in analogy to the lasing threshold in a lossy optical cavity. In particular, when Gilbert damping dominates over spin pumping (which is the case in sufficiently thick magnetic films), i.e., $\alpha\gg\alpha'$, we obtain $\mu'\approx(\alpha/\alpha')\hbar\Omega$, which reproduces the classical Slonczewski's spin-transfer torque instability.\cite{slonczewskiJMMM96}

\subsubsection{Bose-Einstein condensation}

\begin{figure}[pt]
\includegraphics[width=0.7\linewidth,clip=]{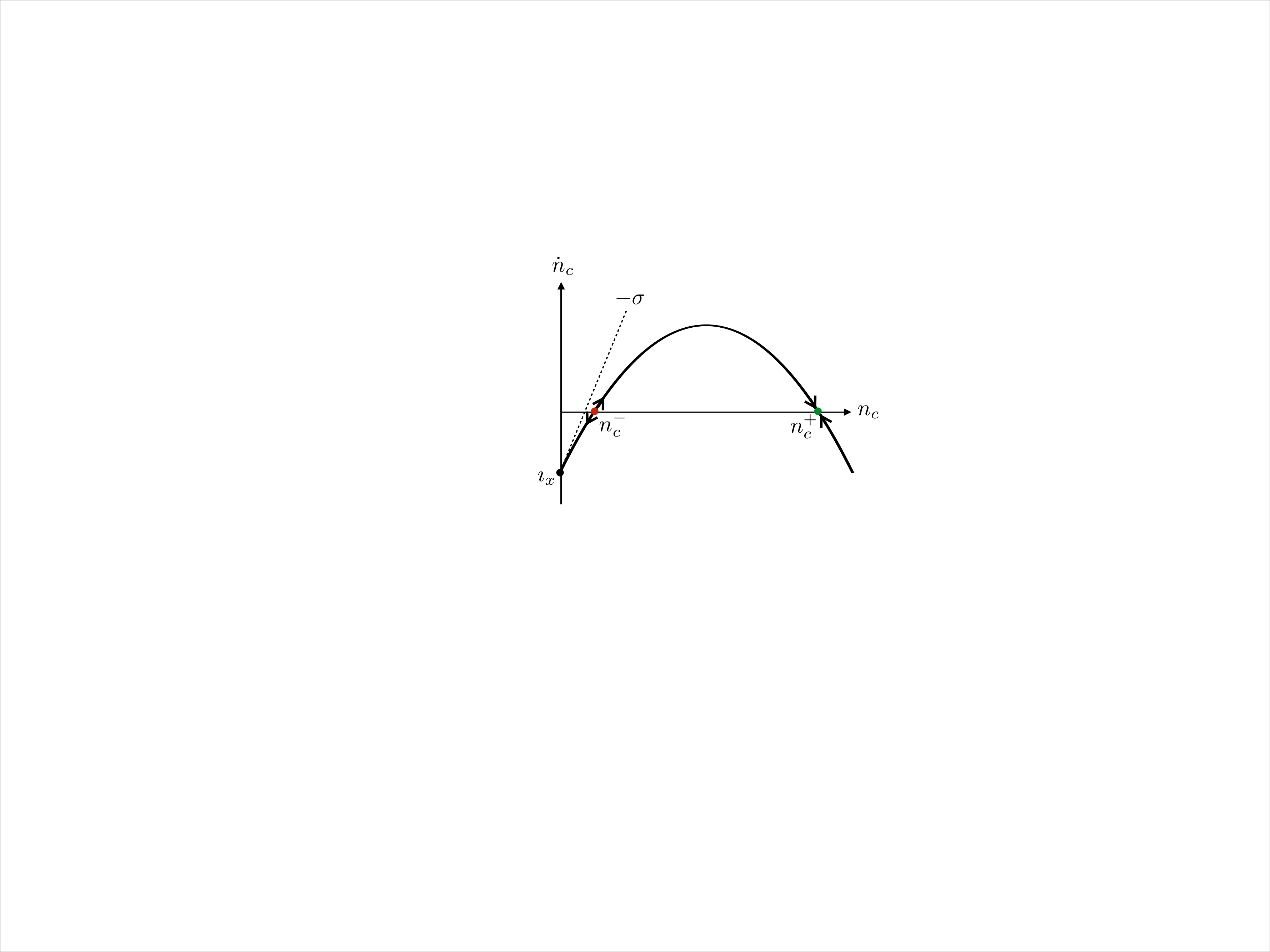}
\caption{A graphical representation for obtaining solutions \eqref{anal} to the equation $\hbar\dot{n}_c=i$ with $i$ given by Eq.~\eqref{ia}. Here, $\sigma<0$ and $\imath_x<0$ (corresponding to region IV$_1$, as described in the text), resulting in two fixed points: unstable at $n_c^-$ and stable at $n_x^+$.}
\label{pb}
\end{figure}

We now focus on the finite-temperature steady-state behavior (fixed points), determined by the condition $i=0$. Namely, for a given set of parameters ($T$, $T'$, $\mu'$, \dots), we look for possible solutions for both NP (defined by the existence of a real value of $\mu^*<0$ for which $i_x+i_x'=0$) and CP (defined by the existence of a real, positive value of $n_c$ for which $i=0$).  While the NP solutions $i_x+i_x'=0$ are found numerically, the analytic expansion in Eq.~(\ref{ia}) allows for a general CP solution to $i=0$:
\begin{equation}
\frac{n_c^{\pm}}{s}=\frac{\pm\sqrt{\sigma^2+4\zeta\imath_x}-\sigma}{2 \zeta}\,,
\label{anal}
\end{equation}
which is depicted in Fig.~\ref{pb}.

\begin{figure}[pt]
\includegraphics[width=0.75\linewidth,clip=]{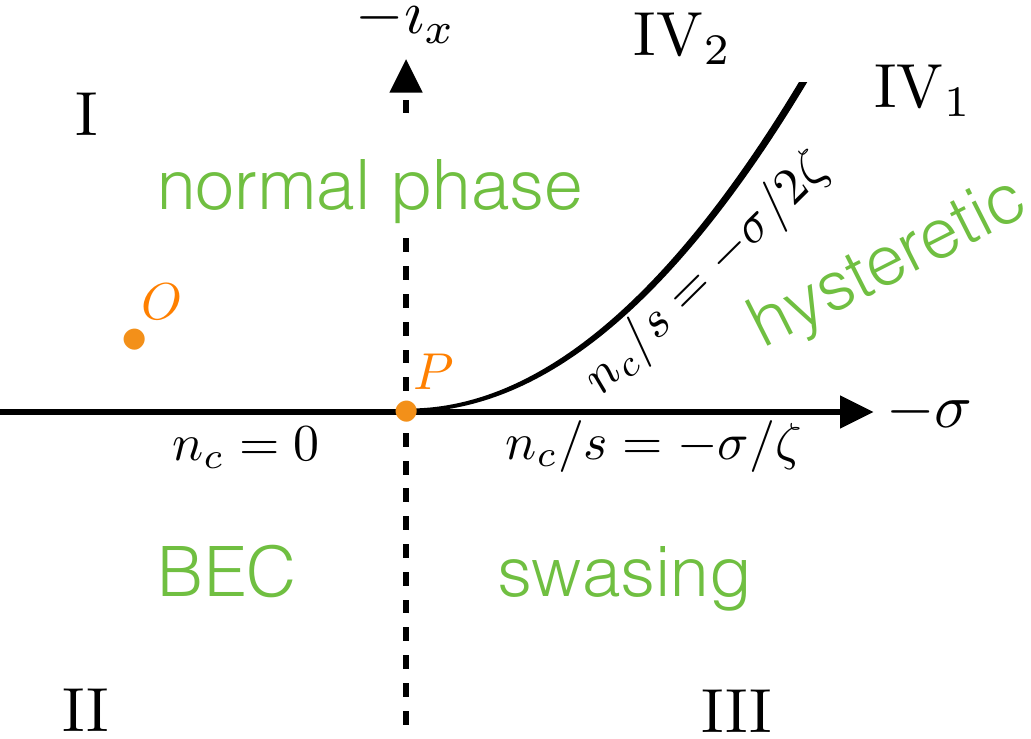}
\caption{Phase diagram for the solutions of Eqs.~\eqref{dnc} and \eqref{ia} for $n_c$ in the abstract $(\sigma,\imath_x)$ space. $O$ stands for the unperturbed (i.e., thermal-equilibrium) point, while $P$ is the critical point for a driven system. The solid lines, $\imath_x=0$ and $\imath_x=-\sigma^2/4\zeta$, trace out phase transitions between distinct dynamic states: second-order transition between the NP and BEC (I/II boundary) and hysteretic first-order transitions at the IV$_2$/IV$_1$ and IV$_1$/III boundaries, where the normalized condensate density, $n_c/s$, jumps by $-\sigma/2\zeta$ and $-\sigma/\zeta$ relative to $0$, respectively. The condensate associated with these first-order transitions is interpreted to be ``swasing."\cite{bergerPRB96}}
\label{pd}
\end{figure}

The resultant phase diagram may be divided into four regions, I-IV, according to the signs of the coefficients $\sigma$ and $\imath_x$:  $\sigma>0$ and $\imath_x<0$ (region I), $\sigma>0$ and $\imath_x>0$ (region II), $\sigma<0$ and $\imath_x>0$ (region III), and $\sigma<0$ and $\imath_x<0$ (region IV). In parameter space, regions I and III each share phase boundaries with regions II and IV. All four regions meet when $\sigma=0$ and $\imath_x=0$, which appears as a single critical point $P$ in the phase diagram. We now discuss in some detail the physical behavior in each of the four regions, with the help of Fig.~\ref{pd} as visual guidance.

In region I, neither solution $n_c^{\pm}$ is real and positive, dictating that the magnons must settle in NP at some $\mu^*<0$ for which $i_x=0$, as we find numerically. In region II, $n_c^+$ represents a real-valued, stable solution to the condensate equation of motion. While the condensate is damped through the second and third terms in Eq.~\eqref{ia}, it is replenished by the thermal cloud, $\imath_x>0$, which can be driven by thermal gradient $T'-T$. The magnet reaches a steady state, wherein angular momentum is pumped into the thermal cloud and transferred to the condensate by magnon-magnon interactions, which in turn decays by the combination of Gilbert damping and spin pumping. Numerically, we find no NP solution coexisting with CP in region II. Note that here $\lim_{\zeta\rightarrow 0}n_c^+/s=\imath_x/\sigma$ is finite even in the absence of the nonlinearity $\zeta$.

The boundary between regions I and II is defined by the condition $\imath_x=0$, corresponding to $n_c=0$. It thus follows that $n_c$ is continuous at the associated NP/CP phase transition, given by $n_c\equiv0$ in region I and $n_c\propto\imath_x$ in the incipient region II. Conversely, $\mu^*\equiv0$ in region II and decreases continuously, $\mu^*<0$, in region I. We identify this dynamic second-order phase transition as a \textit{Bose-Einstein condensation,} whose order parameter is given by $\psi=\sqrt{n_c}e^{i\phi}$, where $\dot{\phi}\approx\omega$. In contrast to swasing, where $\sigma<0$, the condensate decay is compensated here by the thermal magnon injection, $\imath_x>0$, that replenishes it.

\subsubsection{Full phase diagram}

Similarly to region II, region III produces a positive, stable solution $n_c^{+}$ to the condensate equation of motion. In contrast to region II, however, $n_c^{+}/s\to|\sigma|/\zeta$ diverges as $\zeta\rightarrow0$, demonstrating the importance of the nonlinearity $\zeta$ in stemming the condensate growth. In this region, swasing is supplemented with thermal spin transfer $\imath_x$, which increases $n_c^+$. Because no solution to $i_x=0$ exists for $\mu^*<0$ (in our numerical calculation), we conclude that only CP is present in region III. 

Region IV may itself be divided further into two subregions: IV$_1$ and IV$_2$ defined respectively by $\sigma^2\gtrless-4 \zeta\imath_x$. In subregion IV$_1$, both $n_c^+$ and $n_c^-$ are real, but only the former solution is stable (see Fig.~\ref{pb}). Depending on whether $n_c\gtrless n_c^-$ at $t=0$, the magnetic system flows towards CP fixed point at $n_c^+$ or NP, respectively, at $t\to\infty$, indicating CP/NP hysteresis. In contrast, both $n_c^{+}$ and $n_c^{-}$ are complex in subregion IV$_2$, precluding CP. In all of region IV, therefore, an NP solution $\mu^*<0$ to $i_x=0$ exists, which evolves continuously within this region. The CP solution existing in subregion IV$_1$, on the other hand, evolves continuously into a CP swasing phase in region III. Region IV is opposite to II both in the reversal of the sign of $\sigma$ (such that the condensate tends to swase) and $\imath_x$ (such that the thermal magnons are pumped out of the magnet, thus suppressing the condensate). The balancing act between negative $\sigma$ and $\imath_x$, as depicted in Fig.~\ref{pb}, allows for a stable condensate in subregion IV$_2$.

\begin{figure}[pt]
\includegraphics[width=\linewidth,clip=]{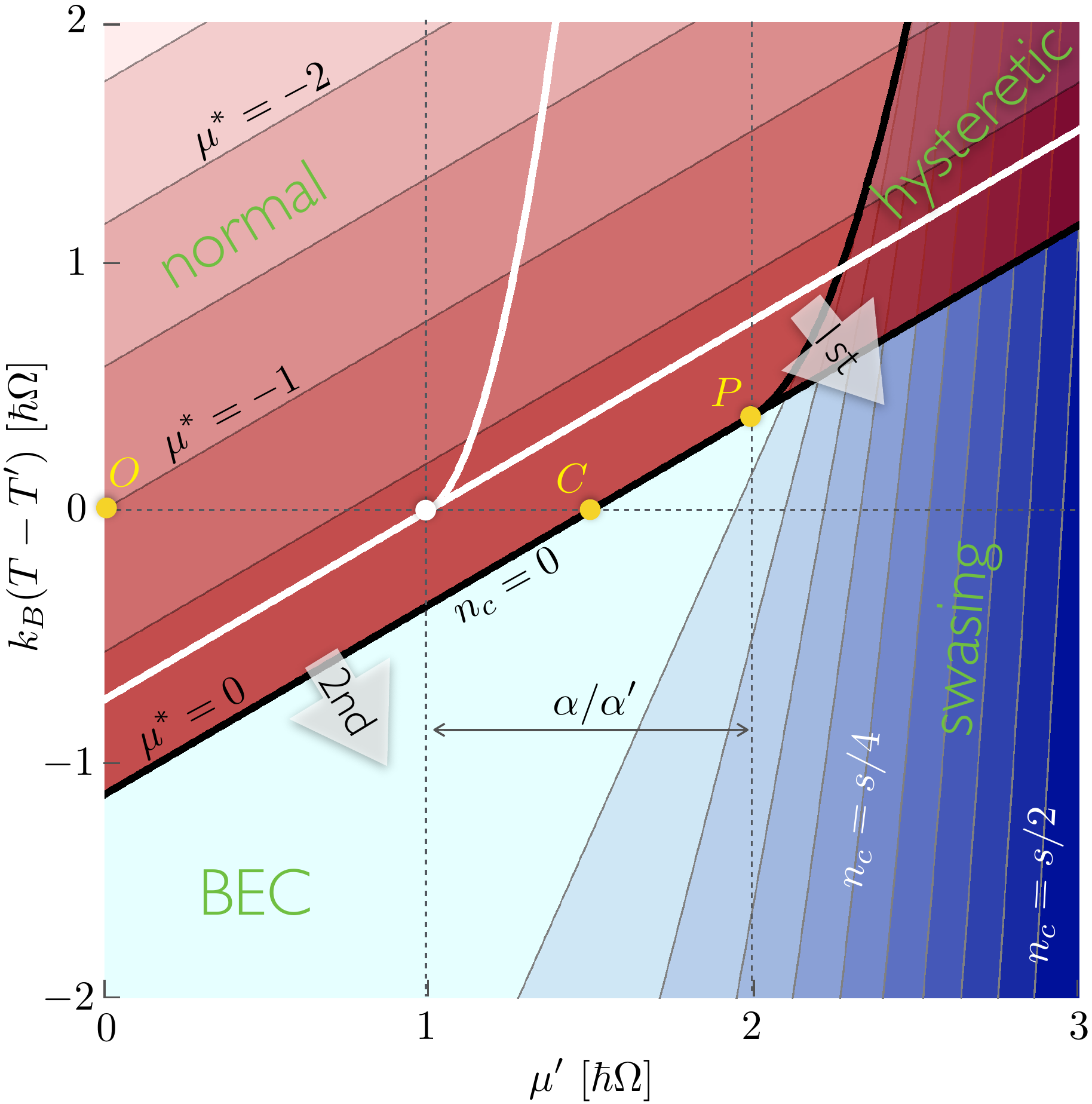}
\caption{Physical phase diagram in the presence of anisotropy $K=\hbar\Omega$ at $k_B T=k_B T''=10^2\hbar\Omega$, $s \left( A/\hbar \Omega \right)^{3/2}=10^4$, and $\alpha/\alpha'=1$ (black curves), calculated using the linearized current $\imath_x$ in Eq.~(\ref{ia}) [see discussion preceding Eq.~(\ref{to})]. The white curves show the idealized $\alpha/\alpha'=0$ case. The analytically evaluated diagram shown here is essentially indistinguishable from the numerical diagram (not shown) produced by the exact expression for $i$ in Eq.~(\ref{ia}). The phase-transition lines and crossovers that delineate different dynamic regimes can be inferred from Fig.~\ref{pd}. } 
\label{fix}
\end{figure}

We summarize the above discussion in Fig.~\ref{pd}: The boundary between regions I and II describes a continuous phase transition between an NP and the Bose-Einstein condensate (BEC). The boundary between I and IV$_2$ is a crossover within the NP, while the boundary between II and III is a crossover between swasing and BEC (both instances of a CP). Boundaries delineating the hysteretic region IV$_1$ define history-dependent first-order transitions: An NP in IV$_1$ jumps to a finite condensate density $n_c/s=-\sigma/\zeta>0$ when entering III, and a CP phase in IV$_1$ jumps from a finite condensate density $n_c/s=-\sigma/2\zeta$ to a normal state with a finite $\mu^*<0$ when entering IV$_2$. All the phase-transition lines and crossovers emanate from the critical point $P$.

When drawing the physical phase diagram in terms of the experimentally-controlled parameters $(\mu',T')$ (which, in turn, determine $\sigma$ and $\imath_x$), the essential structure of Fig.~\ref{pd} is preserved, albeit somewhat distorted, as shown in Fig.~\ref{fix}. While $\mu'$ corresponds linearly to $-\sigma$, according to Eq.~\eqref{s}, $\imath_x$ is generally a nonlinear function of $T'$ and $\mu'$. Since, for a fixed $\mu'$, $\imath_x$ increases with increasing temperature $T'$, however, we can think of $-\imath_x$ as parametrizing $1/T'$ (keeping $T$ fixed). This explains why the structure of the physical phase diagram in Fig.~\ref{fix} is anticipated by Fig.~\ref{pd}.

Let us now parametrize in detail the phase-transition lines depicted in Fig.~\ref{fix}. We denote by $T'_1$ the phase boundary corresponding to the $\imath_x=0$ abscissa in Fig.~\ref{pd} (i.e., the curve delineating phases II and III) and by $T'_2$ the boundary between regions IV$_1$ and IV$_2$ [strictly above the swasing instability \eqref{sw}, i.e., $\mu'/\hbar\Omega> 1+\alpha/\alpha'$], which emanates out of the critical point $P$. When $k_B \left(T-T' \right), \mu', \hbar \Omega \ll k_B T$ (i.e., the ambient temperature sets the largest relevant energy scale), the current $\imath_x$ may be linearized in $k_B \left(T-T' \right)$, $\mu'$, and $\hbar \Omega$, allowing us to analytically derive the expressions for $T'_1 \left(\mu' \right)$  and  $T'_2 \left(\mu' \right)$.  In this regime, the former is linear in $\mu'$ and given by:
\begin{equation}
k_B \left( T-T'_{1} \right)=\frac{2 \zeta_{3/2}}{5\zeta_{5/2}}\left[\mu'-\left(1+ \frac{\alpha}{2 \alpha'}\right)\hbar\Omega\right]\,,
\label{to}
\end{equation}
where $\zeta$ is the Riemann zeta function. Below the swasing threshold, condensate forms when $T' $ exceeds $T'_1$. In the absence of a temperature bias, $T'=T$, Eq.~(\ref{to}) indicates the formation of a condensate when $\mu' $ exceeds $\left(1+\alpha/2\alpha' \right)\hbar\Omega$ (denoted in Fig.~\ref{fix} by $C$).\cite{Note}

\begin{figure}[pt]
\includegraphics[width=\linewidth,clip=]{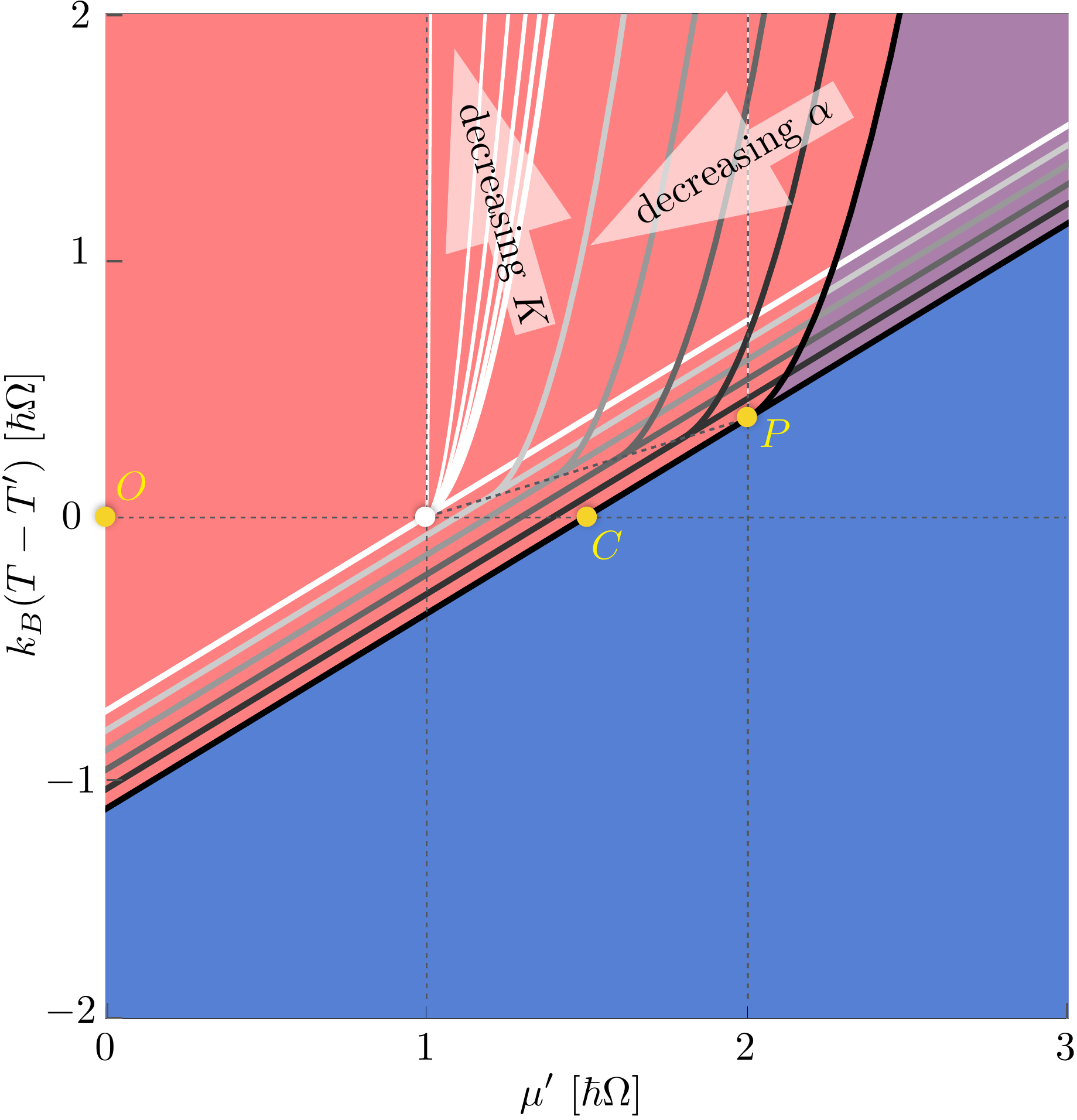}
\caption{Effects of intrinsic damping $\alpha/\alpha'$ (starting at $1$ and decreasing to $0$ in increments of $0.2$) and nonlinearity $K/\hbar\Omega$ (going from $1$ to $0$ in increments of $0.2$), while keeping $\hbar\Omega$ fixed, on the phase-diagram structure, using Eqs.~(\ref{to}) and~(\ref{tt}). Decreasing Gilbert damping $\alpha$ [which lowers the swasing threshold \eqref{sw}] increases the size of the condensate regions, while decreasing anisotropy $K$ increases the size of the hysteretic region [as is evident from Eq.~\eqref{tt}].}
\label{Q}
\end{figure}

The curve $T'_{2}$, in turn, is defined by: 
\begin{equation}
k_B \left( T'_1-T'_2 \right)=\frac{\left[\mu'-(1+\alpha/\alpha' )\hbar\Omega \right]^{2}}{5\Gamma_{5/2}\zeta_{5/2}(1+\alpha/\alpha')}\frac{\pi^2sA^{3/2}}{K( k_B T)^{3/2}}\,,
\label{tt}
\end{equation}
where $\Gamma$ is the gamma function. The curves $T'_1$ and $T'_2$, according to Eqs.~(\ref{to}) and (\ref{tt}) are shown in Fig.~\ref{fix} as solid black lines for $\alpha/\alpha'=1$ and solid white lines $\alpha/\alpha'=0$. The dependence of the transition lines on a gradual change in the strength of damping $\alpha$ and nonlinearity $K$ is shown in Fig.~\ref{Q}.

\subsection{Floating magnon temperature}
\label{free}

In addition to angular momentum, energy transfer from the ferromagnet into the adjacent normal metals and its crystal lattice, in general, also needs to be balanced. In the previous section, we made a simplifying assumption that the magnon temperature was pinned by phonons and/or electrons, which provided a very efficient energy sink. Here we relax that assumption, which necessitates keeping track of the total magnon energy on par with the magnon number. We still, however,  suppose that magnon-magnon interactions are sufficiently strong that the magnons remain internally thermalized to a Bose-Einstein distribution with a well-defined effective temperature $T$ and chemical potential $\mu^*$ (relative to the magnon-band bottom) at all times. We also retain the assumption that cloud and condensate always remain in mutual equilibrium, namely, that $n_c$ vanishes for $\mu^*<0$ (NP) and $n_c>0$ requires that $\mu^*=0$ (CP). In analogy with the condensate and normal-phase spin currents discussed above, we define the condensate and normal-phase energy currents $\jmath_c\equiv\omega(i_c+i_c')$ and $\jmath_x\equiv(j_x+j_x')|_{\mu^*=0,n_c=0}$, respectively, according to Eq.~\eqref{ecrate}, which simplifies the stability analysis of the CP.

At any time, there now exist two dynamical variables.  In NP, these are $\mu^*$ and $T$, governed by the implicit, coupled rate equations $\hbar \dot{n}_x=i_x+i_x'$ and $\dot{e}_x=j_x+j_x'$; in CP,  $n_c$ and $T$, governed by Eqs.~\eqref{totsc} and \eqref{ecrate}. In contrast to the expansion of the magnon current $i$ in $n_c$, Eq.~(\ref{ia}), a simple general analytic expansion of the currents in $T$ in either phase is not possible, and we must resort to a numerical treatment. 

\begin{figure}[pt]
\includegraphics[width=\linewidth,clip=]{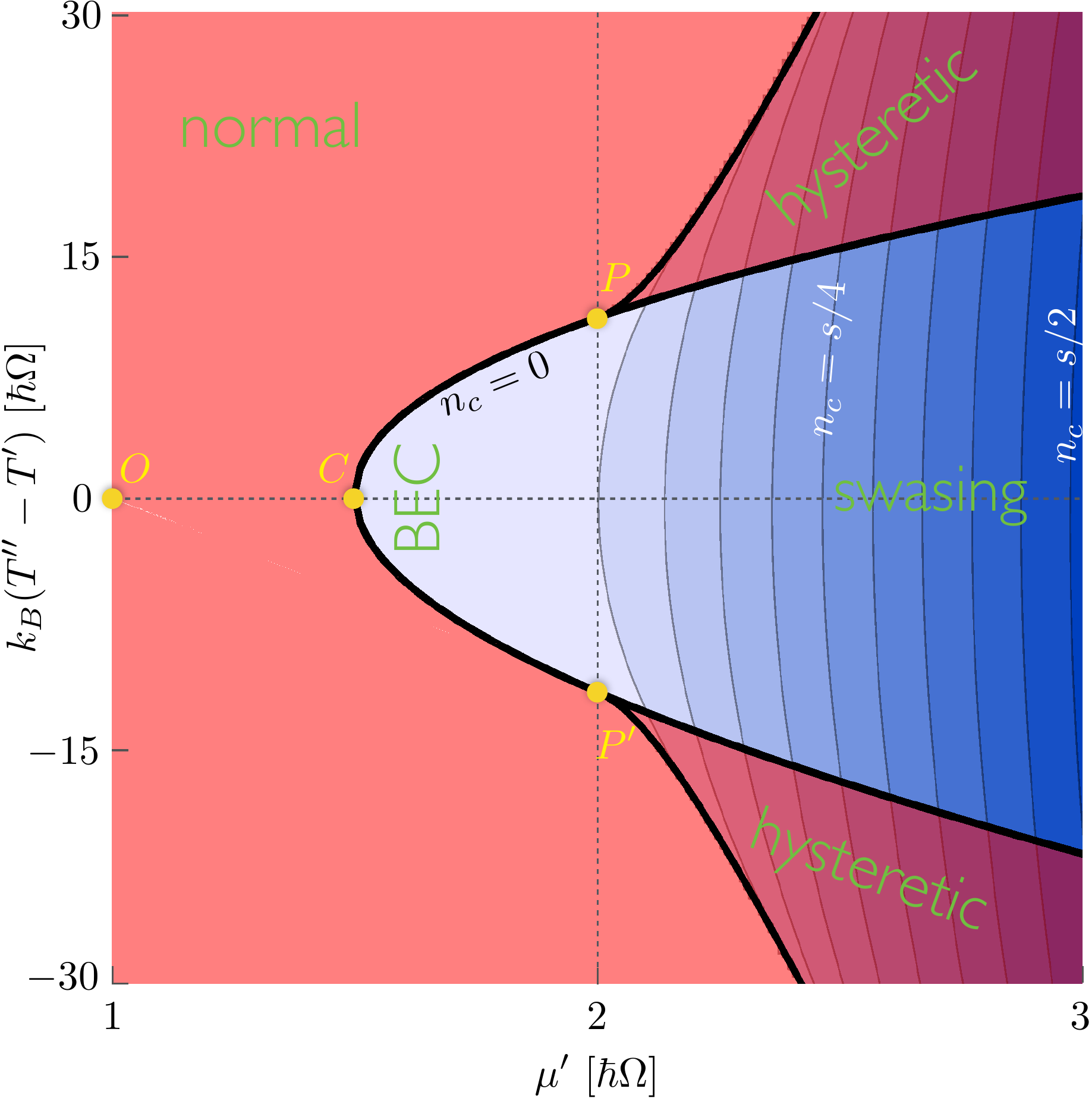}
\caption{Phase diagram with a floating magnon temperature $T$ and density determined by the conditions $i=0$ and $j=0$. Here, $\alpha/\alpha'=1$, $K=\hbar \Omega$, $s \left( A/\hbar \Omega \right)^{3/2}=10^4$, and $k_BT''=10^2\hbar \Omega$, similarly to the other plots.}
\label{Floating}
\end{figure}

In general, the steady-state temperature $T$ and the chemical potential $\mu^*$ (or condensate spin density $n_c$) in each phase are determined from the stable fixed points of the respective pair of coupled rate equations.  The resultant numerical phase diagram is shown in Fig.~\ref{Floating}. The energy accompanying angular-momentum transfer into the ferromagnet creates additional heating (cooling) when $T'\gtrless T''$ (phonon temperature), which hinders (facilitates) condensation relative to the fixed-temperature regime. In particular, condensation via temperature gradient alone (i.e., $\mu'=0$) no longer occurs. Below the swasing instability (i.e., $\sigma>0$), each steady-state solution $\mu^*=0$ in NP coincides with a solution $n_c=0$ in CP, indicating the second-order phase transition. Above the threshold for swasing, on the other hand, hysteretic regions appear, where, depending on the initial conditions, the solutions flow toward stable fixed points in NP or CP. These features are qualitatively similar to those discussed in the case of a fixed magnon temperature, Sec.~\ref{fmt}.

In the case of a low temperature gradient ($T'\approx T''$), the incipient condensation may be understood by expanding $\imath_x$ and $\jmath_x$ in $k_B \left(T''-T' \right)$, $k_B \left(T''-T\right)$, and $\mu'$ (all of which are assumed to be much smaller than the ambient temperature) and solving the steady-state equations to obtain analytic solutions for $T$ and $n_c$.  When $\mu' \leq(1+\alpha/2\alpha')\hbar\Omega$ (denoted by $C$ in Fig.~\ref{Floating}), no condensate solution exists;  increasing $\mu'$ beyond this critical point, the condensate density continuously increases from zero. (Note that the same bias $\mu'$ at $C$ describes the onset of condensation under zero temperature bias both in the fixed- and floating-temperature regimes.) As in the fixed-temperature case, furthermore, when $\mu' \geq(1+\alpha/\alpha')\hbar\Omega$, unstable analytic solutions for $n_c$ appear, suggesting the presence of hysteresis when the temperature gradient is restored; correspondingly, (two) critical points $P$ and $P'$ manifest under a sufficient temperature bias at the swasing instability \eqref{sw}.

The above linearized treatment for the currents $\imath_x$ and $\jmath_x$, however, fails to capture the detailed phase behavior when $T' \neq T''$. There, the spin and energy fluxes that are quadratic in thermal bias are essential for generating the full structure of the phase boundaries depicted in Fig.~\ref{Floating}. In particular, we see that the condensate is suppressed under large temperature biases of both signs: When $T'\ll T''$, the magnons injected by the normal metal are relatively cold but there are ultimately too few of them to precipitate a condensate; when $T'\gg T''$, on the other hand, the magnon injection rate is high but they are too hot to condense. Only at intermediate thermal biases do we reach a compromise between the magnon injection rate and the energy they carry, which allows for a stable condensate to form.

\section{Summary and Outlook}
\label{so}

We studied the steady-state behavior of an insulating magnet driven by the combination of a thermal gradient and spin-transfer torque across its interface with an adjacent normal metal. Agitated by the interfacial magnon-electron and bulk magnon-lattice interactions, our theory describes the emergent nonlinear coherent motion of the condensate (in a quasiequilibrium with the thermal cloud of magnons), demonstrating a surprisingly rich dynamic phase diagram.

The stability analysis of the driven coherent motion depends crucially on the form of the magnetic anisotropy. Our detailed analysis was specific to an easy-plane magnetic film subjected to a large out-of-plane magnetic field (such that the magnetic ground state is nondegenerate). In the case of other geometries and magnetic anisotropies, the phase diagram can be altered. Furthermore, in other configurations, where spin-rotational symmetry is broken in all directions, three-magnon scattering processes would violate magnon conservation, which is built into our model. We nevertheless expect that the essential nature of the first- and second-order instabilities predicted in our model to be generic, although the details would depend on the specific experimental realization. We emphasize that one of the key features predicted by our theory is a possibility of a continuous formation of the condensate in the presence of a temperature gradient alone, which may be less sensitive to the particular magnetic orientation than the more familiar instabilities invoked by a spin-transfer torque.

The presence of coherently-precessing magnetic phases may manifest experimentally in a variety of ways. Collective magnetic modes driven by dc currents may be detected either by their microwave signatures or differential dc response (both in the charge and thermal sectors) in the steady state, similarly to the conventional spin-transfer torque instabilities.\cite{kiselevNAT03} The thermal properties of the magnon condensates, in particular, may differ dramatically from the normal phase, if, for example, the lateral propagation of heat in the plane of our heterostructure can be carried collectively by magnetic dynamics. In addition, unlike thermal magnons that generally travel diffusively with a microscopic spin-diffusion length, low-frequency condensates can carry spin signals over macroscopic distances.\cite{soninAP10,*takeiCM13} Such collective and nonlocal transport signatures of condensation warrant further studies, both theoretically and experimentally.

\acknowledgments

This work was supported in part by FAME (an SRC STARnet center sponsored by MARCO and DARPA), the NSF under Grant No.~DMR-0840965, and by the Kavli Institute for Theoretical Physics through NSF Grant No.~PHY11-25915.


\begin{thebibliography}{25}
\bibitem [{Note()}]{Note}%
  \BibitemOpen
  \bibinfo {note} {The additional factor of $2$ on the right-hand side of Eq.~\eqref{inp} relative to Eq.~\eqref{icp} stems from the fact that the F$\mid$N interface, where the torques develop, coincides with the antinode of the thermal-magnon normal modes.\cite{hoffmanPRB13,*kapelrudPRL13} As a consequence, point $C$ in Fig.~\ref{fix}, which corresponds to $\mu'=(1+\alpha/2\alpha')\hbar\Omega$, precedes the swasing threshold $P$ [at $\mu'=(1+\alpha/\alpha')\hbar\Omega$] along the $\mu'$ axis.}\BibitemShut {Stop}%
\end{thebibliography}
\end{document}